\documentclass{aa}  
\usepackage{graphicx}
\usepackage{txfonts}
\begin{document}
   \title{}

   \subtitle{INTEGRAL high energy detection of the transient IGR~J11321$-$5311}

   \author{V. Sguera\inst{1}, A. Bazzano\inst{2},  A. J. Bird\inst{1}, A. B. Hill\inst{1}, A. J. Dean\inst{1}, L. Bassani\inst{3}, A. Malizia\inst{3}, P. Ubertini\inst{2}.
          }

   \offprints{sguera@astro.soton.ac.uk}
   \institute{School of Physics and Astronomy, University of Southampton, Highfield, SO17 1BJ, UK \and 
   INAF/IASF Roma, via Fosso del Cavaliere 100, 00133 Roma, Italy \and 
   INAF/IASF Bologna, via Piero Gobetti 101, I-40129 Bologna, Italy
         }

   \date{Received 8 March 2007 / accepted 14 April 2007}

  \abstract
   {The transient hard X-ray source IGR J11321$-$5311 was discovered by INTEGRAL on June 2005, during observations of the
    Crux spiral arm. To date, this is the only detection of the source to be reported by any X/$\gamma$-ray mission.}
   {To characterize  the behaviour and hence  the nature of the source through temporal and spectral IBIS analysis.}
   {Detailed spectral and temporal analysis has been performed using standard INTEGRAL software OSA v.5.1.}
   {To date, IGR J11321$-$5311 has been detected only once. It was active for $\sim$ 3.5 hours, a short and bright flare  lasting  $\sim$1.5 hours 
     is evident in the IBIS light curve. It reached a peak flux of  $\sim$ 80 mCrab or 2.2$\times$10$^{-9}$ erg cm$^{-2}$ s$^{-1}$ (20--300 keV), 
      corresponding to a peak luminosity  of $\sim$ 1.1$\times$10$^{37}$ erg s$^{-1}$ (assuming a distance of 6.5 kpc).
     During the outburst,  the source was detected with a significance of $\sim$ 18$\sigma$ (20--300 keV) and $\sim$ 8$\sigma$ (100--300 keV).
      The spectrum of the total outburst activity (17--300 keV) is best fitted by the sum of a power law ($\Gamma$=0.55$\pm$0.18) 
      plus a black body  (kT=1.0$^{+0.2}_{-0.3}$ keV), with no evidence for a break up to 300 keV. 
      A spectral analysis at Science Window level revealed an evident hardening of the spectrum through the outburst.
      The IBIS data were searched for pulsations with no positive result.}
   {The X-ray spectral shape and the flaring behaviour favour the hypothesis that IGR~J11321$-$5311 is an Anomalous X-ray Pulsar, 
    though  a different nature can not be firmly rejected at the present stage.}


   \maketitle

\section{Introduction}
The IBIS instrument (Ubertini et al. 2003), on board the INTEGRAL satellite (Winkler et al. 2003),
is playing a key role in detecting many transient hard X-ray sources, thanks to its large field of view, good sensitivity and spatial resolution.
The new hard X-ray transient IGR~J11321$-$5311 was discovered by IBIS  on June 2005 during a deep observation of the
Crux spiral Arm (Krivonos et al. 2005). It was active for a few hours with 
average fluxes of $\sim$ 30 mCrab (17--60 keV) and $\sim$ 90 mCrab (60--200 keV). Subsequently, the flux diminished below 
$\sim$ 3 mCrab (17--60 keV). So far, the INTEGRAL detection is the only one to be reported and no further
informations  or study at different wavelengths 
are available. Here we report for the first time
on detailed spectral and timing analysis of the IBIS data for  IGR~J11321$-$5311, aimed at understanding the origin of its hard X-ray emission.
\section{INTEGRAL results} 
The reduction and analysis of the ISGRI data have been performed 
using the INTEGRAL Offline Scientific Analysis (OSA) v.5.1.
INTEGRAL observations are typically divided into short pointings (Science Windows, ScWs) 
of $\sim$ 2000 seconds duration. Throughout  the paper, uncertainties are given at a 90\% confidence level.

\begin{figure}[t!]
\centering
\includegraphics[width=8cm,height=4cm]{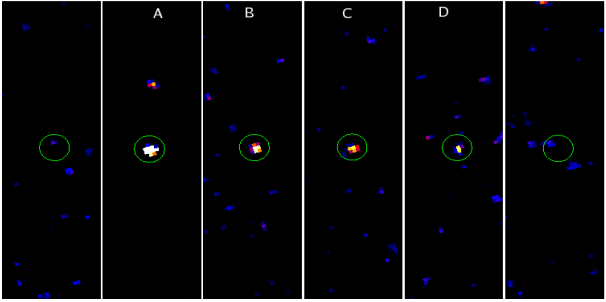}
\caption{ISGRI ScWs significance image sequence (20--300 keV) of the outburst from IGR~J11321$-$5311 
detected on June 2005.
The source (encircled) was not detected in the first ScW, then it was detected during the next 4 consecutive ScWs with a significance, from left to right, 
equal to 12$\sigma$,  8$\sigma$, 5$\sigma$ and 4$\sigma$, respectively.
Finally, in the last ScW the source was not detected.
\label{fig:IGR11321_sig_seq.ps}}
\end{figure}

IGR~J11321$-$5311 was detected by IBIS in 4 consecutive ScWs
in the energy range 20--300 keV. Figure ~\ref{fig:IGR11321_sig_seq.ps}  shows a sequence of significance maps around these ScWs.
The source was not detected in the first ScW, then it was detected during the next 4 consecutive ScWs with a significance, 
from left to right, equal to 12$\sigma$,  8$\sigma$, 5$\sigma$ and 4$\sigma$, respectively.
Finally, in the last ScW the source was undetectable.
Summing the 4 ScWs in a  mosaic, IGR~J11321$-$5311 is detected at $\sim$ 18$\sigma$ level in the energy band 
20--300 keV and $\sim$ 8$\sigma$ in 100--300 keV. 
The position of IGR~J11321$-$5311 as taken from the mosaic is RA=11 32 15.72, Dec=-53 11 41, error radius=1$^{'}$.5 (90\% confidence).
The source is located off the Galactic plane (b=7$^{\circ}$.85), in the direction of the Crux spiral arm tangent. We can assume an approximate distance to
the source of $\sim$ 6.5 kpc.

\begin{figure}[t!]
\centering
\includegraphics[width=6cm,height=8cm,angle=270]{7439fig2.ps}
\caption{ISGRI light curve (20--300 keV)   of the outburst from IGR~J11321$-$5311 on June 2005. 
\label{fig:IGR11321_LC.ps}}
\centering
\includegraphics[width=7cm,height=6cm,angle=270]{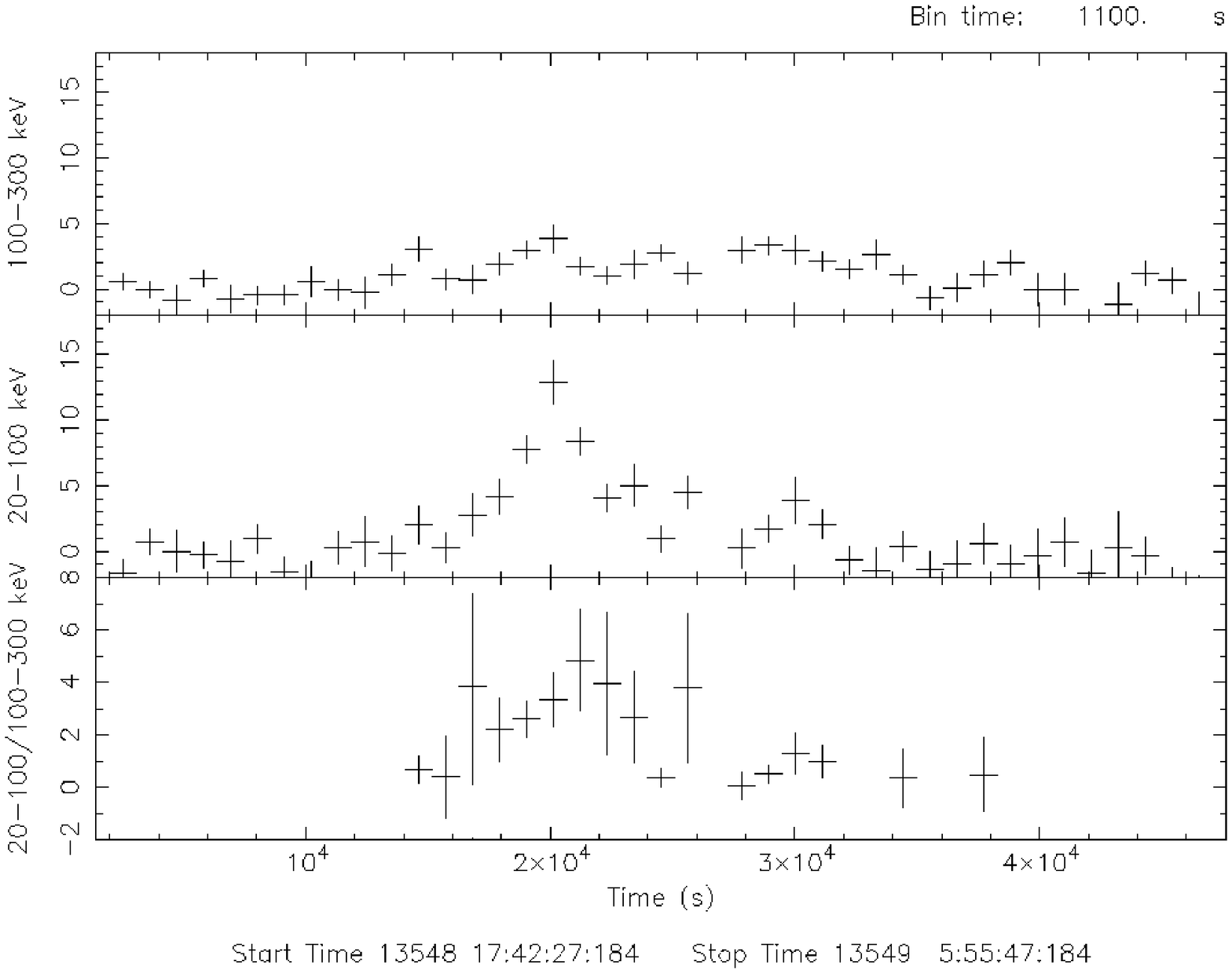}
\caption{ISGRI light curves of IGR~J11321$-$5311 in the energy bands 20--100 and  100--300 keV. In the lower panel, the ratio of the two light curves is shown.
\label{fig:LC_20_100_300_HR.ps}}
\end{figure}

Figure ~\ref{fig:IGR11321_LC.ps} shows the 20--300 keV ISGRI light curve of the outburst shown in figure  ~\ref{fig:IGR11321_sig_seq.ps}.
We assume the beginning of the first ScW during which the source was detected as being the start time of the outburst and similarly the burst
stop time to be the end of the last ScW during which the source was detected.
As we can note from the light curve, initially  the source is not detected, the 2$\sigma$ upper limit (ScW level) is $\sim$ 11 mCrab or 3$\times$10$^{-10}$ 
erg cm$^{-2}$ s$^{-1}$ (20--300 keV). Then it became detectable at 21:59:49 UTC (27 June 2005)
reaching in only $\sim$ 1 hour a peak flux of $\sim$ 80 mCrab or $\sim$ 2.2$\times$10$^{-9}$ erg cm$^{-2}$ s$^{-1}$ (20--300 keV).
For the assumed distance, the corresponding peak luminosity is  $\sim$ 1.1$\times$10$^{37}$ erg s$^{-1}$ (20--300 keV).
Subsequently, the flux gradually decreased and then the source became undetectable on 28 June 2005 at 01:22:33 UTC, with the same upper limit as above.
The duration of the total outburst activity was $\sim$ 3.5 hours, during which the flare lasted $\sim$ 1.5 hours.
Figure ~\ref{fig:LC_20_100_300_HR.ps} shows the light curve in  two different energy bands (20--100 keV and 100--300 keV) and their ratio 20--100/100--300 keV.
The latter suggests a possible spectral evolution of the source during the outburst.

In order to search for pulsations, a time resolved ISGRI light curve with barycentric correction (1 second binning) was
produced in the 20--100 keV band, using the OSA ii-light tool. However no clear signal was detected using  the Lomb-Scargle periodogram method.
Using a Monte-Carlo approach, it was possible to assess the data
quality of the fine timing light curve and our sensitivity
to periodic signals.  Simulated light curves were generated
which had the same sampling and statistical properties of
the data but which had a sinusoidal modulation of $\sim$10
seconds introduced.  The results of the Monte-Carlo
simulations found that to detect 100\% pulsed emission a
signal strength $\sim$3 times stronger than that of the data was
required; 50\% pulsed emission required a signal $\sim$9 times
stronger.  The implication of the Monte-Carlo results is
that the statistical quality of the ISGRI data is
insufficient to detect pulsations, and we cannot rule out their presence.

\begin{figure}[t!]
\centering
\includegraphics[width=5cm,height=7cm,angle=270]{7439fig4.ps}
\caption{Unfolded power law plus black body spectrum (17--300 keV) of IGR~J11321$-$5311 extracted during its total outburst activity.
\label{fig:IGR11321_spectrum.ps}}
\centering
\includegraphics[width=5cm,height=7cm,angle=270]{7439fig5.ps}
\caption{Unfolded power law plus black body spectra (17--300 keV) of IGR~J11321$-$5311 extracted from the sum of ScW A and B (top), C and D (bottom). 
\label{fig:spectrum_scw_42_43_44_45.ps}}
\end{figure}
\begin{figure}[t!]
\centering
\includegraphics[width=6cm,height=4cm]{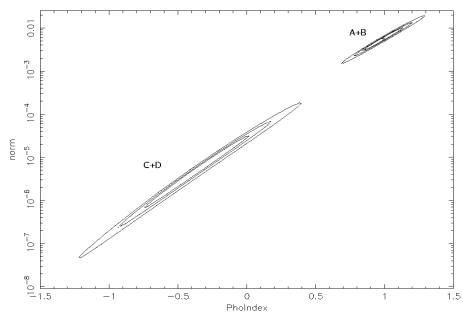}
\caption{Confidence contour levels for the photon index and the normalization constant of the power law model of
the spectra extracted from the sum of ScW A+B and C+D.
\label{fig:cont.ps}}
\end{figure}

The spectrum  (17--300 keV)  extracted from the sum of the four ScWs (A,B,C and D in figure ~\ref{fig:IGR11321_sig_seq.ps}) 
during which the source was detected  cannot be fitted by a single power law ($\chi^{2}_{\nu}$=4.7, d.o.f.48). 
The presence of an additional component below $\sim$ 25 keV is clearly evident. The fit did not improve by adding 
another power law component ($\chi^{2}_{\nu}$=4.95, d.o.f.46), whereas it improved significantly by adding a black body  
($\chi^{2}_{\nu}$=0.8, d.o.f.46); the inclusion of this component was found to be significant at a confidence level greater than 99.99\% using a F-test.
In this case the best fit parameters are $\Gamma$=0.55$\pm$0.18 and kT=1.0$^{+0.2}_{-0.3}$ keV.
The unfolded power law plus black body spectrum is displayed in Figure ~\ref{fig:IGR11321_spectrum.ps}.
The spectrum does not show any break  up to $\sim$ 300 keV. For the assumed distance, 
a radius of the emitting black body region equal to R=7.7 km  can be  inferred, which is compatible with the typical value (10 km) of a neutron star.
The flux of the black body component is $\sim$ 3.8$\times$10$^{-10}$ erg cm$^{-2}$ s$^{-1}$, accounting for $\sim$ 14\% of the total flux.
IGR~J11321$-$5311 was unfortunately outside the narrower JEM--X FOV so that it is not possible to add any information at lower energies. 
A spectral analysis at ScW level has also been performed, searching for a possible spectral evolution during the outburst, as highlighted in Figure 3.
Spectra extracted from ScW A and B (see Figure ~\ref{fig:IGR11321_sig_seq.ps}) are very similar, both being best fitted by a 
power law plus a black body model (17--300 keV) with almost 
identical  values of the best fit parameters. It was therefore  reasonable to extract a spectrum from the sum of the two ScWs A+B to get a better model fit to the data. 
In this case, the best fit parameters of the power law plus black body best fit ($\chi^{2}_{\nu}$=0.8, d.o.f.46) are $\Gamma$=1$\pm$0.17  
and kT=0.9$^{+0.44}_{-0.25}$ keV. The best fit parameters concerning the spectrum extracted from the sum of ScW C+D (see Figure ~\ref{fig:IGR11321_sig_seq.ps}) 
are  $\Gamma$=--0.36$^{+0.18}_{-0.44}$  and kT=1.0$^{+0.3}_{-0.3}$ keV ($\chi^{2}_{\nu}$=0.55, d.o.f.46). The spectrum of IGR~J11321$-$5311
during the final part of its outburst activity is much harder than that of the beginning of the outburst.
This can be clearly noted in Figure 5, which shows the two spectra extracted from the sum of ScW  A+B and C+D, and in Figure 6 which displays 
their confidence contour levels for the photon index and the normalization constant of the power law.

IGR~J11321$-$5311 was not reported in the third IBIS catalog (Bird et al. 2007) although the region of the sky including the source was observed for a 
total of  $\sim$ 1.15 Ms. It is worth pointing out that the third IBIS catalog input dataset did not include the INTEGRAL data pertaining 
the detection of the source discussed here, since they were not public at the time. 
It is therefore possible to infer a 2$\sigma$ IBIS upper limit to the quiescent flux of IGR~J11321$-$5311, 
equal to $\sim$ 0.5 mCrab or 3.4$\times$10$^{-12}$ erg cm$^{-2}$ s$^{-1}$ (20--40 keV).
Assuming again a distance of $\sim$ 6.5 kpc, the corresponding  upper limit on the quiescent luminosity 
is $\sim$ 1.5$\times$10$^{34}$ ergs$^{-1}$ (20--40 keV).
Recently, {\emph Swift} XRT  performed two targeted observations  of IGR~J11321$-$5311 (exposure time of $\sim$ 2.5 and 3.3 ks), but no X-ray objects were  
detected inside the ISGRI error box (Landi et al. 2007). Moreover, the source has been targeted several times by the RXTE satellite, 
but detections were never reported. We can then conclude that IGR~J11321$-$5311 spend a considerable fraction of its time in quiescence.

\section{Discussion}

We reported on spectral and timing analysis of the ISGRI detection of IGR~J11321$-$5311  on June 2005, to date the only one
reported in the literature.
The source  was active for only $\sim$ 3.5 hours, being characterized by a short flare ($\sim$ 1.5 hours duration).
The  17--300 keV spectrum of the total outburst activity was best fitted by a hard power law ($\Gamma$=0.55$\pm$0.18) plus a black body model
(kT=1.0$^{+0.2}_{-0.3}$ keV). There is no indication in the spectrum of any break up to 300 keV.
Spectral analysis at ScW level revealed spectral evolution of the source during the outburst, its spectrum 
being much harder ($\Gamma$$\sim$--0.36) at the end than at the  beginning ($\Gamma$$\sim$1).

No X-ray, gamma-ray  and radio sources are  located  inside the ISGRI error circle (radius $\sim$ 1$^{'}$.5) 
in any of the  available  catalogs in the HEASARC database. As for the optical and infrared band, 
the ISGRI error circle is  naturally too large for  a fruitful  identification using catalogs such as USNO--B1.0 or 2MASS.
A much more refined source position is strongly needed to this aim. It is worth pointing out that Negueruela \& Schurch (2006) 
used existing photometric catalogues to search for early type stars within the error circles of unidentified X-ray source believed
to be HMXBs. Their method was very efficient at finding reddened OB stars, resulting in the detection of the counterpart to several 
unidentified X-ray sources. However, no suitable candidates were found in the field of IGR~J11321$-$5311 (Negueruela \& Schurch 2006).

Though the observational data on  IGR~J11321$-$5311 are few, all the 
findings reported here may give indications on its nature. 
The temporal and the spectral behaviour of the source  are reminiscent of  
outburst characteristics from two different class of hard X-ray emitters: magnetars and  black hole transients. 

Indeed, IGR~J11321$-$5311 could be a new member of the Anomalous X-ray Pulsars (AXPs) group. 
A review by Kaspi (2006) outlines the recent observational progress on temporal and spectral behaviour of the seven, 
possibly nine AXPs so far detected. They have pulsation periods ranging from 6 to 12 s, large period derivatives 
($\sim$ 10$^{-13}$--10$^{-10}$ ss$^{-1}$), very strong magnetic fields (0.6--7.1$\times$10$^{14}$ G) and some of them are 
located  inside Supernova Remnants.
Their relatively high X-ray luminosity ($\sim$ 10$^{34}$--10$^{36}$ erg s$^{-1}$) cannot be accounted for by rotational energy losses, moreover no 
convincing evidence for a companion star  has been found for any of them. Many observational properties support the idea that AXPs 
are magnetars, isolated neutron stars powered by the decay of their huge magnetic fields. 
Spectra in the traditional X-ray band (0.5--10 keV) are well described by two components; a blackbody  
(kT$\sim$0.3--0.6 keV) due to internal heating caused by the intense magnetic field decay plus a relatively steep power law (2$<$$\Gamma$$<$4) resulting from resonant 
scattering of the thermal seed photons off magnetospheric currents in the twisted magnetoshere. The softness of the spectra predicts
non detections above 10 keV,
however recent INTEGRAL observations have showed that  AXPs are very hard X-ray emitters with 
spectra extending well above 100 keV and characterized by  no break up to 300 keV (Kuiper et al. 2006).
Their pulsed spectra are exceptionally hard with a photon index in the range from -1 to 1, while the photon index of  the total spectra (sum of pulsed and unpulsed components)
are in the range from 1 to 1.4 (Kuiper et al. 2006). Until recently AXPs were believed to be steady X-ray emitters, 
however one of the recent interesting discoveries is the range and diversity  of their X-ray variability properties. 
AXPs show long term flux variations 
and  short term variability such as outbursting behaviour. A short outburst has been fortuitously  detected from 1E 2259+589 (Kaspi  et al. 2003, Woods et al. 2004),
it lasted a few hours and it was accompanied by a dramatic hardening of the spectrum. XTE J1810$-$197 (Woods et al. 2005) 
and CXOU J164710.2$-$455216 (Israel et al. 2007) are 
other examples of recently discovered transient and bursting AXPs.
The flaring behaviour,  the spectral  characteristics with no break up to 300 keV and hardening of the spectrum, 
the peak and quiescent luminosities of  IGR~J11321$-$5311 ($\sim$ 10$^{37}$ erg s$^{-1}$ and upper limit of 
$\sim$ 10$^{34}$ erg s$^{-1}$ respectively), all favour the hypothes of an AXP.  
Being typically $|$b$|$$<$1$^{\circ}$ for the AXPs sample,  the IGR~J11321$-$5311 location off the Galactic plane (b=7$^{\circ}$.85) 
could be suggestive of a close by AXP. However if we assume a closer distance, i.e. 3 kpc, the upper limit 
on the quiescent luminosity would drop to 10$^{33}$ erg s$^{-1}$, which is lower when compared with the average values for the known AXPs.

An alternative to the AXP scenario is a black hole (BH) transient nature.
Here we briefly indicate a  few BHs showing short and peculiar outbursts: SAX J1819.3$-$2525, XTE J0421+560 
and Cyg X$-$1. The X-ray transient SAX~J1819.3$-$2525 (see in't  Zand et al. 2000 and Revnivtsev et al. 2002 for a review)
showed  a series of very bright and short  X-ray flares 
occurred within less than 1.5--2 days. The brightest one reached a level of $\sim$ 12 Crab (2--10 keV) and then the source 
totally disappeared within 0.3 day. Spectra accumulated during the peak  
are reminiscent of a black hole when in  the low state with $\Gamma$$<$2, cut-off at energy 100--200 keV.
XTE J0421+560/CI Cam 
was discovered  in 1998 by RXTE (Smith et al. 1998), it brightened very quickly up to 2 Crab a few hours after the discovery 
and decayed exponentially with a very short e-folding time (0.6 day). The 20--100 
keV BATSE data on the decline are consistent with a power law with a photon 
index of --3.9 while there is marginal evidence of a harder
spectrum during the rise and/or peak (Belloni et al. 1999). In the case of Cyg X-1,  seven episodes of 
strong hard X-ray emission occurred within  9 years (Golenetskii et al., 2003). These outbursts have duration up to $\sim$8 hours 
and reached peak fluxes of 3$\times$10$^{-7}$ erg cm$^{-2}$ s$^{-1}$ (15--300 keV). All the spectral characteristics of the above black hole outburst 
examples do not show the peculiarity of IGR J11321--5311, i.e. flat spectral index and lack of any spectral break up to 300 keV. 
Moreover, the duration of the outburst detected by INTEGRAL from IGR~J11321$-$5311 ($\sim$ 3.5 hours) is very short when  compared 
to typical transient activity of accreting black hole binaries, which is of the order of months.

The location of IGR~J11321$-$5311  off the Galactic plane (b=7$^{\circ}$.85) could suggest also an AGN nature.
However, BL Lacs and Blazars are known to display flares with timescales of days or weeks, significantly longer than the flare detected from IGR~J11321$-$5311
($\sim$ 5500 seconds). A possible very short hard X-ray flare ($\sim$ 2000 seconds duration) was reported only in one case, the Blazar 
NRAO 530 (Foschini et al. 2006). Moreover, Blazar or BL Lac are known to be strong radio sources, but  no radio source is reported inside the ISGRI 
error circle of IGR~J11321$-$5311 using all the radio catalogues available in the HEASARC database. 
However it is worth pointing out that the region of the sky containing  IGR~J11321$-$5311 has not yet been covered 
by deep radio surveys (e.g. NVSS).

Further detections of  IGR~J11321$-$5311 by INTEGRAL or other X-ray missions, as well as multiwavelength observations,
could be very useful to shed more light on this enigmatic  hard X-ray transient source.

\begin{acknowledgements}
We thank the anonymous referee for useful comments which helped us to improve the paper.
This research has been supported by University of Southampton School of Physics
and Astronomy. AB,LB,AM,PU acknowledge founding support via contract  ASI/INAF I/023/05/0

\end{acknowledgements}


\begin{thebibliography}{}

  


\bibitem[]{}  Bird, A.J., Malizia, A., Bazzano, A., et al. 2007, in press, astro-ph/0611493

\bibitem[]{}  Belloni, T., Dieters S., Van Den Ancker, et al. 1999, ApJ, 527, 345 
		
\bibitem[]{} 	Foschini, L.; Pian, E.; Maraschi, L.,et al. 2006, A\&A,450,77


\bibitem[]{}  	Golenetskii, S.; Aptekar, R.; Frederiks, D., et al. 2003, ApJ,596,1113


\bibitem[]{} in't  Zand  J. et al 2000, A\&A,357,520
	
\bibitem[]{} Israel, G. L.; Campana, S.; Dall'Osso, S., et al. 2007, in press, astro-ph/0703684

\bibitem[]{}  Landi, R., Kennea J., Giommi, P., et al. 2007, ATEL 991


	
\bibitem[]{} Negueruela, I.\& Schurch, M. P. E., 2007, A\&A, 461, 631

	

\bibitem[]{} Kaspi, V. M.; Gavriil, F. P.; Woods,, et al. 2003,  ApJ,588L,93
\bibitem[]{} Kaspi, 2006, astro-ph/0610304

\bibitem[]{}  Krivonos R., Molkov S., Revnivtsev M., et al. 2005, ATEL 545


	
\bibitem[]{} Kuiper, L.; Hermsen, W.; den Hartog, P. R.;, et al. 2006, ApJ, 645, 556

	
\bibitem[]{}	Revnivtsev, M.; Gilfanov, M.; Churazov, E., et al. 2002,A\&A,391,1013

\bibitem[]{} Smith et al., 1998,IAUC 6855


\bibitem[]{}  Ubertini P., Lebrun, F., Di Cocco, G., et al. 2003, A\&A, 411, L131

	
\bibitem[]{} 	Winkler,C.; Courvoisier,T.; Di Cocco, G., et al. 2003, A\&A, 411, 1

		
\bibitem[]{}	Woods, P.M.; Kouveliotou, C.; Gavriil,  P.; et al. 2005, ApJ, 629, 985
\bibitem[]{}	Woods, P. M.; Kaspi, V. M.; Thompson, C., et al. 2004, AIP Conference Proceedings, Vol. 714, 298





\end{thebibliography}
\end{document}